# Anomalous valley Hall effect in antiferromagnetic monolayers

Wenhui Du, Rui Peng, Zhonglin He, Ying Dai*, Baibiao Huang, Yandong Ma*

School of Physics, State Key Laboratory of Crystal Materials, Shandong University, Shandanan Str. 27, Jinan 250100, China

*Corresponding author: daiy60@sina.com (Y.D.); yandong.ma@sdu.edu.cn (Y.M.)

## Abstract

Anomalous valley Hall (AVH) effect is a fundamental transport phenomenon in the field of condensed-matter physics. Usually, the research on AVH effect is mainly focused on 2D lattices with ferromagnetic order. Here, by means of model analysis, we present a general design principle for realizing AVH effect in antiferromagnetic monolayers, which involves the introduction of nonequilibrium potentials to break of PT symmetry. Using first-principles calculations, we further demonstrate this design principle by stacking antiferromagnetic monolayer $MnPSe_3$ on ferroelectric monolayer $Sc_2CO_2$ and achieve the AVH effect. The AVH effect can be well controlled by modulating the stacking pattern. In addition, by reversing the ferroelectric polarization of $Sc_2CO_2$ via electric field, the AVH effect in monolayer $MnPSe_3$ can be readily switched on or off. The underlying physics are revealed in detail. Our findings open up a new direction of research on exploring AVH effect.

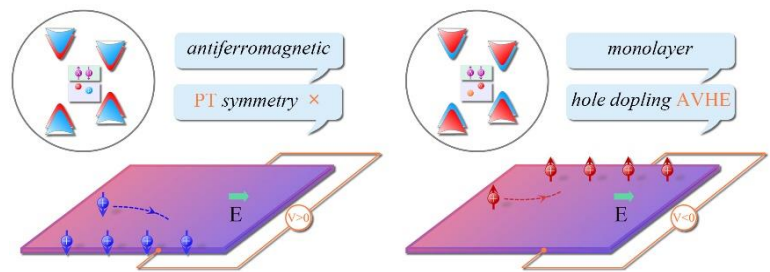

# I. INTRODUCTION

Valley, characterizing energy extrema of conduction or valence band, is an emerging degree of freedom of carriers in condensed-matter materials [1,2]. Benefiting from the large separation in momentum space, the valley degree of freedom is particularly stable against low-energy phonons and smooth deformations [3-5]. Analogous to charge and spin, the valley degree can be utilized to encode information and perform logic operations, leading to the concept of valleytronics [6,7]. Early interest in valley dates back to the works in 1970s on silicon inversion layer [8,9]. In recent years, following the discoveries of intrinsic physical properties associated with valley occupancy in two-dimensional (2D) lattices [10,11], rapid experimental and theoretical progress [12-46] has been made in the field of valleytronics at both the fundamental and applied levels.

As a new piece of puzzle in the Hall family, the anomalous valley Hall (AVH) effect in valley-polarized materials lies at the heart of valleytronics. In principle, polarized light is able to induce AVH effect in valley-polarized monolayers, however, it is a dynamic process and subjected to the life time of carriers, which is not applicable for practical valleytronics [27,28]. Physically, there are two essential ingredients for realizing stable AVH effect: one is magnetism and the other is spin-orbital coupling (SOC). The former can be either intrinsic magnetism [29-36] or extrinsic magnetism caused by magnetic doping [37-40], magnetic field [41,42] and magnetic proximity effect [43-46], while the latter is related to the chemical compositions. The past years have seen impressive progress for identifying such AVH effect in single-layer materials [29-46]. In all the prior studies, there is one default assumption, namely, the magnetism must be ferromagnetic. To our knowledge, it is still unclear how to realize stable AVH effect in single-layer materials with an antiferromagnetic order.

In the present work, we show by model analysis that the realization of stable AVH effect can be extended to single-layer materials with an antiferromagnetic order. The proposed general design principle for this extension correlates with the introduction of nonequilibrium potentials to break the PT symmetry. Based on first-principles calculations, we further demonstrate this design principle by stacking antiferromagnetic monolayer $MnPSe_3$ on ferroelectric monolayer $Sc_2CO_2$ and realize the AVH effect. Such AVH effect is shown to exhibit a tantalizing stacking pattern depended character. Moreover, we reveal that the AVH effect in monolayer $MnPSe_3$ can be switched on or off by reversing the ferroelectric polarization of monolayer $Sc_2CO_2$. This extension of AVH effect to single-layer materials with an antiferromagnetic order is of great significance from both fundamental perspective and for potential use in devices.

# II. METHODS

Our first-principles calculations are performed based on density functional theory (DFT) methods as implemented in the Vienna ab initio simulation package (VASP) [47,48]. The generalized gradient approximation (GGA) in form of Predew-Burke-Ernzerhof (PBE) functional is used to describe the exchange-correlation interaction [49]. The cutoff energy is set to 500 eV. Crystal structures are fully relaxed with the convergence criteria of $10^{-5}$ eV and 0.01 eV/Å for energy and force, respectively. The Brillouin zone is sampled with Monkhorst–Pack grids of 9×9×1. To avoid interactions between adjacent periodical structures, the vacuum space along the z direction is set to 30 Å. The zero damping DFT-D3 method is utilized to treat the van der Waals (vdW) interaction [50]. To describe the strong correlation effects, the effective on-site Hubbard term of U = 4 eV is set for the 3d electrons of the Mn atom, as this value is employed in previous works [44,51,52]. Berry curvature is calculated using the maximally localized Wannier function method as implemented in the WANNIER90 package [53].

## III. RESULTS AND DISCUSSION

Our proposed design principle for realizing AVH effect in single-layer materials with an antiferromagnetic order is schematically illustrated in **Fig. 1**. Without losing the generality, we take monolayer $MnPSe_3$ as an example to discuss the scheme by considering the fact that monolayer $MnPSe_3$ is known as a typical Néel antiferromagnetic semiconductor with spontaneous valley polarization. Monolayer $MnPSe_3$ exhibits a hexagonal lattice with the space group of $D_{3d}$. Without considering exchange interaction, its structure hosts the inversion symmetry P. In monolayer $MnPSe_3$, the Mn atoms constitute two sublattices, which are referred to as A and B sublattices. Although the A and B sublattices are structurally equivalent, their spin orientations are opposite. In this regard, neither inversion symmetry P nor time-reversal symmetry T is preserved. However, it shows invariance under the simultaneous time reversal and spatial inversion, namely, PT symmetry. Because of the PT symmetry, the spin and valley polarization occur spontaneously in such antiferromagnetic monolayers, as schematically shown in **Fig. 1(a)**. Also protected by the PT symmetry, the valley spin splitting is prohibited, yielding the spin degeneracy for the K and K′ valleys. Such spin degeneracy in antiferromagnetic monolayers forbids the realization of AVH effect. For example, upon shifting Fermi level between the K and K′ valleys in the conduction bands, the spin-up and spin-down electrons from the K′ valley would accumulate at opposite edges of the sample in the presence of an in-plane electric field, preventing the AVH effect.

If we introduce nonequilibrium potentials to neighboring Mn atoms, the equivalence of the A and B sublattices is deformed, which will break the PT symmetry. Without the protection of PT symmetry, the valley spin splitting at the K and K′ valleys would be achieved [see **Fig. 1(b)**], rendering the observation of AVH effect. Such valley spin splitting is different from the case in nonmagnetic materials, where the valley spin splitting relates to Zeeman splitting [14,15,54]. To preserve the valley physics, the nonequilibrium potentials introduced here should be regular. More importantly, it is natural to expect that the spin orders at both the K and K′ valleys can be reversed by reversing the nonequilibrium potentials introduced to A and B sublattices [**Fig. 1(c)**], which would enrich the AVH effect in antiferromagnetic monolayers.

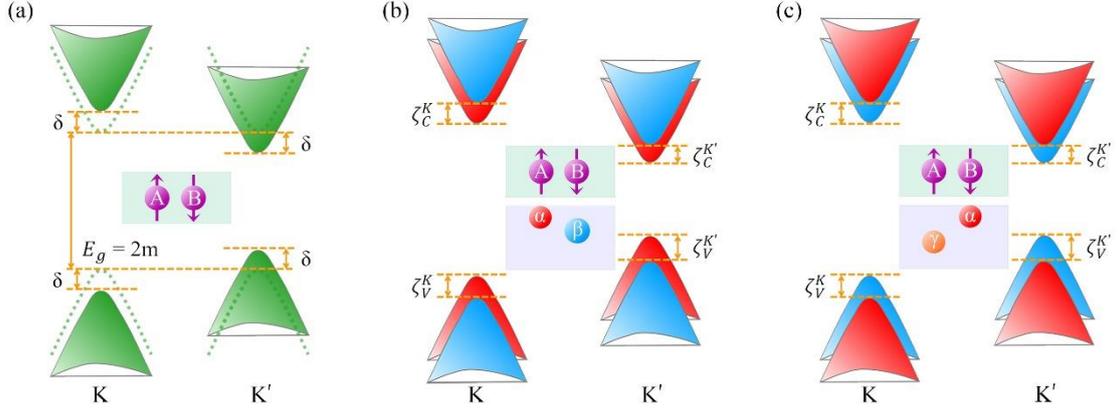

**FIG. 1.** Schematic diagram for the K and K′ valleys of monolayer MnPSe$_3$ (a) without and (b, c) with nonequilibrium potentials on the A and B sublattices. Dotted and solid cones in (a) represent the valley without and with considering SOC, respectively. Red and blue cones in (b, c) correspond to two spin states, respectively.

Given the proposed design principle, we construct a **k·p** model to verify it. The spin-full Hamiltonian with excluding SOC for low-energy quasiparticles near the K(K′) valley of a honeycomb lattice can be expressed as [16]

$$H^0 = v_F s_0 (\tau_z \sigma_x p_x + \tau_0 \sigma_y p_y) + m s_z \tau_0 \sigma_z$$

Here, $p$ and $v_F$ represent the momentum operator and the massless Fermi velocity, respectively. $\tau_\alpha$, $\sigma_\alpha$, and $s_\alpha (\alpha = x, y, z, 0)$ are the Pauli matrices for the valley, isospin, and spin degree of freedom, respectively. The mass term (m), referring to an AFM perturbation, creates a band gap of $E_g = 2m$ for both spins, preserving the spin and energy degeneracy. When taking SOC into account, the Hamiltonian for SOC is introduced as

$$H^{SOC} = \delta s_z \tau_z \sigma_z$$

where $\delta$ is the SOC parameter. In the presence of SOC, the band gap is enlarged by $2\delta$ at the K valley and reduced by $2\delta$ at the K′ valley. As a result, the energetic degeneracy of the K and K′ valleys are lifted, but the spin degeneracy is preserved due to the invariance of the $H^0 + H^{SOC}$ under simultaneous time reversal and spatial inversion. The term for the nonequilibrium potential between two sublattices can be written as:

$$H_{ex} = \frac{\zeta}{2} s_0 \tau_0 \sigma_z$$

where $\frac{\zeta}{2}$ represents the strength of the nonequilibrium potential. In this case, the $H^0 + H^{SOC} + H^{ex}$ would not be invariance under simultaneous time reversal and spatial inversion. This drives the bands from one sublattice to shift upward with respect to those from the other, resulting in a valley spin splitting of $\zeta$ at the K and K′ valleys. According to the total Hamiltonian $H = H^0 + H^{SOC} + H^{ex}$, the eigenvalues of the highest valence bands are estimated to be $(-m - \delta - \frac{\zeta}{2})$ and $(-m - \delta + \frac{\zeta}{2})$ for spin-up and spin-down at the K valley, respectively, and $(-m + \delta - \frac{\zeta}{2})$ and $(-m + \delta + \frac{\zeta}{2})$ for spin-up and spin-down at K′ valley. And the eigenvalues of the lowest conduction bands are found to be $(m + \delta + \frac{\zeta}{2})$ and $(m + \delta - \frac{\zeta}{2})$ for spin-up and spin-down at the K valley, respectively, and $(m - \delta + \frac{\zeta}{2})$ and $(m - \delta - \frac{\zeta}{2})$ for spin-up and spin-down at K′ valley. When reversing the nonequilibrium potential between two sublattices, the sign of $H_{ex}$ is reversed, leading to the opposite valley spin splitting at the K and K′ valleys. These results firmly indicate that the proposed design principle for realizing AVH effect in antiferromagnetic monolayers is feasible physically.

Having established the feasibility of the proposed design principle, next we discuss its realization in real materials. The proximity effect is proposed here to introduce nonequilibrium potentials on the A and B sites regularly. For detail, we consider monolayer $Sc_2CO_2$ as the substrate to induce the proximity effect in monolayer $MnPSe_3$. The crystal structure of monolayer $Sc_2CO_2$ is shown in **Fig. S1(b)** in Ref. [55]. It exhibits a hexagonal lattice with the space group P3m1. Due to the asymmetric displacement of inner C atomic layer with respect to Sc atomic layers, it hosts intrinsic ferroelectricity with out-of-plane polarization [56]. The band structure shown in **Fig. S1(d)** suggests that monolayer $Sc_2CO_2$ is a semiconductor with an indirect band gap of 1.79 eV. Concerning the stacking between monolayer $MnPSe_3$ and $Sc_2CO_2$, a 2 × 2 supercell of $Sc_2CO_2$ is adopted to match the unit cell of $MnPSe_3$. Three typical stacking patterns (i.e., h-I$_\downarrow$, h-II$_\downarrow$ and h-III$_\downarrow$) with the polarization of $Sc_2CO_2$ pointing away from the interface are first considered, as shown in **Fig. 2(a)**. In h-I$_\downarrow$ pattern, the $Mn_1$ atoms sit above the top-O sites of $Sc_2CO_2$, and the $Mn_2$ atoms are right above the Sc atoms in the second layer of $Sc_2CO_2$. In h-II$_\downarrow$ pattern, the $Mn_1$ atoms lie above the C atoms in the middle layer, and the $Mn_2$ atoms are right above the top-O atoms. In h-III$_\downarrow$ pattern, the $Mn_1$ atoms sit above the Sc atoms in the second layer, and $Mn_2$ atoms lie above the C atom in the middle layer of $Sc_2CO_2$.

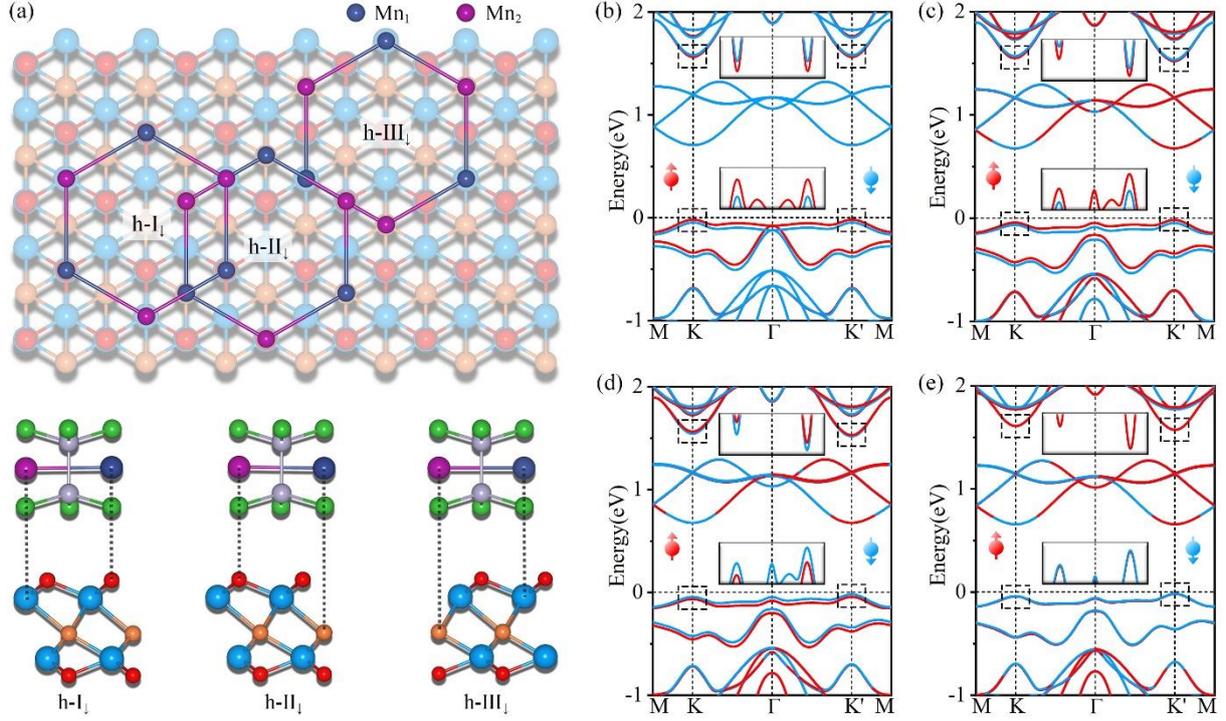

**FIG. 2**. (a) Crystal structures of h-I$_\downarrow$, h-II$_\downarrow$ and h-III$_\downarrow$. Band structures of h-I$_\downarrow$ (b) without and (c) with considering SOC. (d) Band structure of h-II$_\downarrow$ with considering SOC. (e) Band structure of h-III$_\downarrow$ with considering SOC. Blue and red lines in (b-e) correspond to spin-down and spin-up states, respectively. The Fermi level is set to 0 eV.

In h-I$_\downarrow$ configuration, the Mn$_1$ atoms would experience an extra interaction due to their proximity to the top-O sites of Sc$_2$CO$_2$, while the Mn$_2$ atoms would endure with a relatively small interaction as they lie farther. This can be described by the scenario illustrated in **Fig. 1(b)**, which introduces nonequilibrium potentials into the two sublattices of Mn atoms. The band structure of h-I$_\downarrow$ configuration without considering SOC is shown in **Fig. 2(b)** and **S2(a)**. The conduction band minimum (CBM) and valence band maximum (VBM) are contributed by monolayer Sc$_2$CO$_2$ and MnPSe$_3$, respectively, forming a type-II band alignment. The degenerate valleys of MnPSe$_3$ in both the conduction and valence bands locating at the K and K′ points are preserved after stacking on Sc$_2$CO$_2$. However, as shown in **Fig. 2(b)**, the interesting valley spin splitting occurs in both the bottom conduction and valence bands of MnPSe$_3$ at the K/K′ point, which correlates to the nonequilibrium potentials induced by stacking. Although the valleys of MnPSe$_3$ in the conduction bands submerge in the bands of Sc$_2$CO$_2$, they are still investigated for comparison. The valley spin splitting in the valence bands ($\zeta_V^K = \zeta_V^{K'}$) and conduction bands ($\zeta_C^K = \zeta_C^{K'}$) are estimated to be 28.5 and 16.7 meV, respectively. Such sizeable valley spin splitting stands in sharp contrast to the valley spin splitting in nonmagnetic materials which are induced by SOC and P breaking.

When taking SOC into account, as shown in **Fig. 2(c)**, the character of valley spin splitting of MnPSe$_3$ is maintained at both the K and K′ valleys, but the values of valley spin splitting at K and K′ valleys are no longer equal; see **Table S1**. In this case, the energetic degeneracy of the K and K′ valleys in MnPSe$_3$ is lifted, leading to the spontaneous valley polarization. Particularly, the valley polarization in the valence band is as large as 24.6 meV. The coexistence of valley spin splitting and valley polarization would ensure the observation of AVH effect in antiferromagnetic monolayer of MnPSe$_3$, as we will show later.

As discussed above, the valley spin splitting relates to the nonequilibrium potentials induced by the proximity effect. Therefore, by engineering the stacking pattern, the valley spin splitting can be modulated. In h-II$_↓$ configuration, the Mn$_2$ atoms experience an extra interaction due to their proximity to the top-O sites of Sc$_2$CO$_2$, while the Mn$_1$ atoms endure with a relatively small interaction as they lie farther. This corresponds to the scenario illustrated in **Fig. 1(c)**. Accordingly, the nonequilibrium potentials introduced to Mn$_1$ and Mn$_2$ atoms in h-II$_↓$ configuration is reversed with respect to those in h-I$_↓$ configuration. The band structure of h-II$_↓$ configuration with considering SOC is presented in **Fig.2(d)**. It shares similar features to the band structure of h-I$_↓$. The interesting valley spin splitting and spontaneous valley polarization are observed at the K and K′ valleys of MnPSe$_3$, and the magnitudes are comparable to that in h-I$_↓$ configurations. However, the sign of the valley spin splitting is opposite to that of h-I$_↓$ configurations, which is consistent with the proposed design principle. Such simultaneous existence of valley spin splitting and valley polarization would facilitate the AVH effect as well.

Different from the cases of h-I$_↓$ and h-II$_↓$ configurations, both the Mn$_1$ and Mn$_2$ atoms in h-III$_↓$ configuration experience a weak extra interaction since they lie far away from the atoms of Sc$_2$CO$_2$. Thus, the nonequilibrium potentials introduced to Mn$_1$ and Mn$_2$ atoms by the proximity effect would be rather weak. According to the proposed design principle, the valley spin splitting of MnPSe$_3$ in h-III$_↓$ configuration would be tiny. The band structure of h-III$_↓$ configuration with considering SOC is displayed in **Fig. 2(e)**. As expected, the valley spin splitting is as small as 2.3 meV (2.6 meV) at the K (K′) valley in the valence band, while it is even 1 meV (1 meV) in the conduction band of MnPSe$_3$, although large spontaneous valley polarization is preserved. In this case, the observation of AVH effect in h-III$_↓$ configuration is difficult to achieve. Therefore, the valley spin splitting in MnPSe$_3$ is indeed stacking dependent and thus can be engineered by modulating the stacking pattern, while holds great promise for practical applications.

The nonequilibrium potentials introduced to Mn$_1$ and Mn$_2$ atoms by the proximity effect for all these three stacking patterns can also be straightforwardly reflected by the variation of magnetic moments. For freestanding MnPSe$_3$, the magnetic moments on Mn$_1$ and Mn$_2$ atoms are found to be 4.52 and -4.52 μB, respectively, which is protected by the PT symmetry. When stacking MnPSe₃ on Sc₂CO₂ to form the h-I↓ configuration, the magnetic moment on Mn₁ atom changes to 4.508 μB, while the magnetic moment on Mn₂ atom experiences with a relatively slight change (-4.513 μB). This suggests the nonequilibrium potentials on the Mn₁ and Mn₂ atoms. As compared with h-I↓ configuration, the absolute values of the magnetic moments on Mn₁ and Mn₂ atoms in h-II↓ configuration are exchanged, but the signs of the magnetic moments remain the same. Namely, the magnetic moments on Mn₁ and Mn₂ atoms in h-II↓ configuration are found to be 4.513 and -4.508 μB, respectively. This can be easily understood by recalling the reversal of the atomic environments for Mn₁ and Mn₂ atoms in h-I↓ and h-II↓ configurations. For h-III↓ configuration, the magnetic moments on Mn₁ and Mn₂ atoms slightly change to 4.514 and -4.514 μB, respectively. The preservation of identical absolute values of the magnetic moments on Mn₁ and Mn₂ atoms in h-III↓ configuration relates to the weak extra interactions on both Mn atoms.

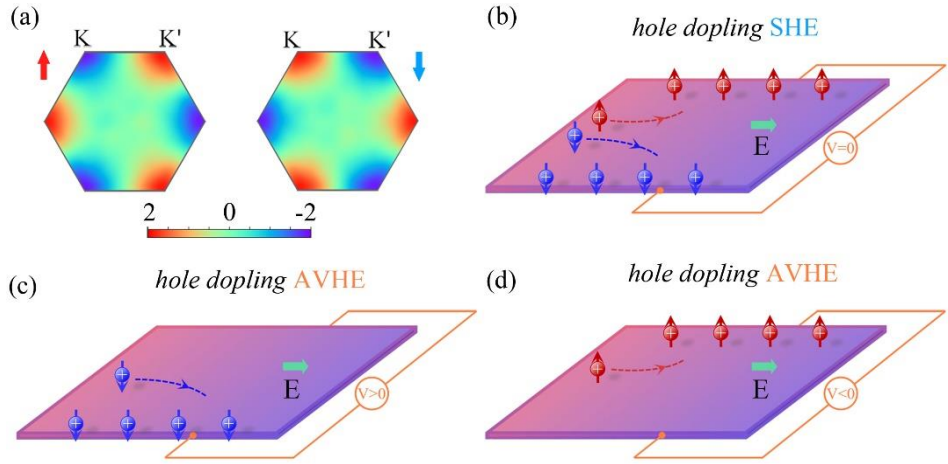

**FIG. 3**. (a) Berry curvatures of the spin-up and spin-down channels over the 2D Brillouin zone for monolayer MnPSe₃. (b) Diagram of the spin and valley Hall effects in monolayer MnPSe₃ under hole doping in presence of an in-plane electric field. Diagrams of the spin Hall and AVH Hall effects in (c) h-I↓ and (d) h-II↓ configurations under hole doping in presence of an in-plane electric field. Red/blue arrows in (b, c, d) indicate spin-up/down states, and the "+"/"-" symbols indicate holes/electrons from the K′ valley.

In 2D hexagonal lattices with inversion symmetry breaking, the K and K′ valleys will exhibit a nonzero Berry curvature along the out-of-plane direction. The Berry curvature is defined as [57]

$$\Omega(k) = -\sum_n \sum_{n \neq n'} f_n \frac{2Im\langle\psi_{nk}|v_x|\psi_{n'k}\rangle\langle\psi_{n'k}|v_y|\psi_{nk}\rangle}{(E_n - E_{n'})^2}$$

Here, $f_n$ is the Fermi-Dirac distribution function, $\psi_{nk}$ is the Bloch wave function with eigenvalue

$E_n$, and $v_x/v_y$ is the velocity operator along x/y direction. The calculated Berry curvatures of the spin-up and spin-down channels over the 2D Brillouin zone for monolayer MnPSe$_3$ are shown in **Fig. 3(a)**. Obviously, the Berry curvatures for the same spin state around the K and K′ valleys are opposite, and the Berry curvatures for spin-up and spin-down states at the same valley are opposite as well. Note that under a longitudinal in-plane electric field, the Bloch carriers will acquire an anomalous transverse velocity proportional to the Berry curvature: $v_a \sim E \times \Omega(k)$. In monolayer MnPSe$_3$, when shifting the Fermi level between the K and K′ valleys in the valence band, protected by the PT symmetry, the spin-down and spin-up holes from the K′ valley will gather to opposite edges of the sample in the presence of an in-plane electric field, see **Fig. 3(b)**. In this regard, a net spin current is generated, which results in the spin Hall effect, however, the AVH effect is absent. In h-I$_↓$ configuration, because of the additional valley spin splitting, the valley degeneracy is lifted. When shifting the Fermi level between the K and K′ valleys in the valence band, only the spin-down holes from the K′ valley move to the bottom boundary of the sample under an in-plane electric field; see **Fig. 3(c)**. This gives rise to the long sought AVH effect in antiferromagnetic monolayers. In addition, the accumulated spin-down holes will result in a net charge/spin current. Similar to the case of h-I$_↓$ configuration, when shifting the Fermi level between the K and K′ valleys in the valence band of the h-II$_↓$ configuration, the spin-up holes from the K′ valley will accumulate at the top edge of the sample [**Fig. 3(d)**], which results into the AVH effect as well.

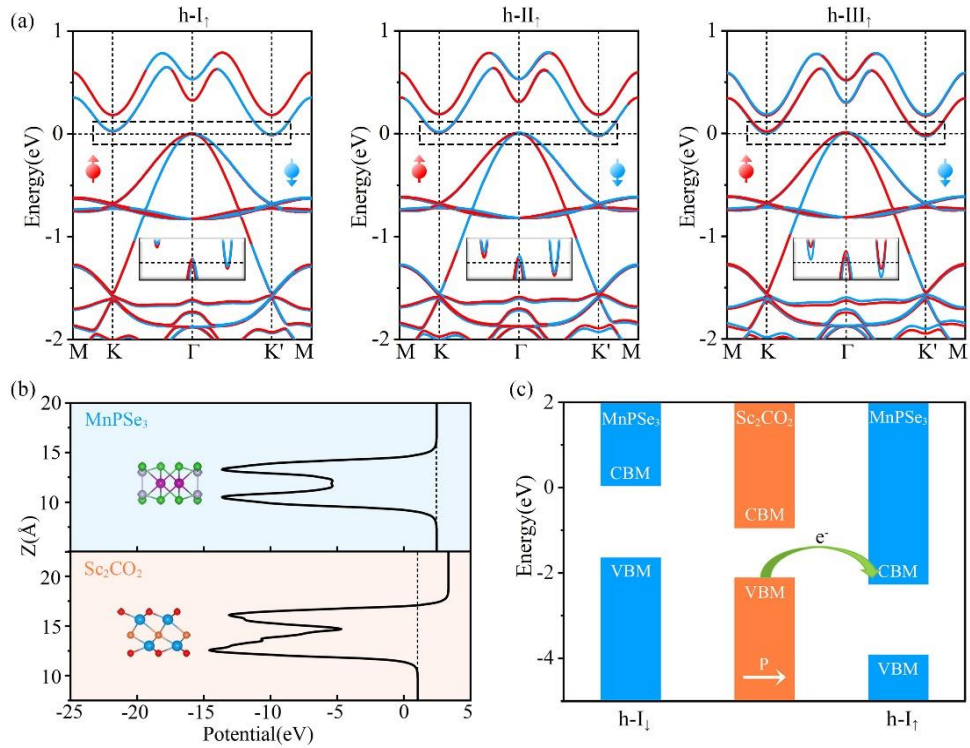

**Fig. 4**. Band structures of h-I$_↑$, h-II$_↑$ and h-III$_↑$ with considering SOC. Blue and red lines in (a) correspond to spin-down and spin-up states, respectively. The Fermi level is set to 0 eV. (b) Plane-

averaged electrostatic potentials of monolayer MnPSe$_3$ and Sc$_2$CO$_2$ along the z direction. (c) Band alignments of h-I$_\downarrow$ and h-I$_\uparrow$ with respect to the vacuum level.

Note that monolayer Sc$_2$CO$_2$ is a 2D ferroelectric crystal with an out-of-plane polarization, its polarization direction can be reversed under a short-term out-of-plane electric field. Such ferroelectric switching might affect the electronic properties of the MnPSe$_3$/Sc$_2$CO$_2$. In the following, we investigate the properties of the three typical stacking patterns with switching the ferroelectric polarization of Sc$_2$CO$_2$ from downward to upward, which are referred to as h-I$_\uparrow$, h-II$_\uparrow$, h-III$_\uparrow$, respectively; see **Fig. S3**. **Fig. 4(a)** presents the band structures of h-I$_\uparrow$, h-II$_\uparrow$ and h-III$_\uparrow$ configurations with considering SOC. Different from the cases with polarization pointing from the interfaces, the band structures of all these three configurations share similar characters. In all these three configurations, the valley physics in the valence bands around the Fermi level is disappeared. More interestingly, the Fermi level crosses the bottom of the conduction band and top of the valence band, yielding a metallic nature. In other word, under the ferroelectric switching, all these three configurations undergo a semiconductor-to-metal transition. Such transition is also accompanied with the disappearance of valley physics as well as the AVH effect. Accordingly, the AVH effect in h-I$_\downarrow$ and h-II$_\downarrow$ configurations is ferroelectric controllable, benefiting for developing controllable valleytronic devices.

To understand the underlying mechanism of the ferroelectric-polarization-dependent properties, we calculate the plane-average electrostatic potentials of monolayer MnPSe$_3$, monolayer Sc$_2$CO$_2$, h-I$_\downarrow$ and h-I$_\uparrow$ [**Fig. 4(b)** and **Fig. S4**]. It can be seen that arising from the asymmetric structure, intrinsic polarization occurs along the out-of-plane direction, leading to the different work functions at the two sides of monolayer Sc$_2$CO$_2$. Accordingly, upon contacting monolayer MnPSe$_3$ with Sc$_2$CO$_2$ in different polarized states, the distinctly different band alignments forms, as shown in **Fig. 4(c)**. When the polarization of monolayer Sc$_2$CO$_2$ points away from the interface, the CBM of Sc$_2$CO$_2$ locates above the VBM of MnPSe$_3$, inhibiting the transfer of electrons between Sc$_2$CO$_2$ and MnPSe$_3$. The valley feature in the valence band from MnPSe$_3$ is preserved. In contrast, when the polarization pointing to the interface, the CBM of MnPSe$_3$ shifts below the VBM of Sc$_2$CO$_2$, prompting the electron transfer from Sc$_2$CO$_2$ to MnPSe$_3$, leading to the metallic nature. The charge transfer character between Sc$_2$CO$_2$ to MnPSe$_3$ in h-I$_\uparrow$, h-II$_\uparrow$ and h-III$_\uparrow$ configurations is also confirmed by the charge density differences shown in **Fig. S5**. While for h-I$_\downarrow$, h-II$_\downarrow$ and h-III$_\downarrow$ configurations, the charge redistribution occurs around the interface. With these results in hand, we can understand the polarization dependent behaviors in MnPSe$_3$/Sc$_2$CO$_2$.

## IV. CONCLUSION

To summarize, through model analysis, a general design principle for realizing stable AVH effect in antiferromagnetic monolayers is proposed, which involves the introduction of nonequilibrium potentials to $Mn_1$ and $Mn_2$ atoms to break the PT symmetry. Using first-principles calculations, the proposed design principle is further demonstrated by stacking antiferromagnetic monolayer $MnPSe_3$ on ferroelectric monolayer $Sc_2CO_2$. The realized AVH effect hosts the stacking pattern depended character. In addition, the AVH effect in $MnPSe_3/Sc_2CO_2$ can be switched on or off under ferroelectric switching of monolayer $Sc_2CO_2$.

## ACKNOWLEDGEMENT

This work is supported by the National Natural Science Foundation of China (Nos. 11804190 and 12074217), Shandong Provincial Natural Science Foundation (Nos. ZR2019QA011 and ZR2019MEM013), Shandong Provincial Key Research and Development Program (Major Scientific and Technological Innovation Project) (No. 2019JZZY010302), Shandong Provincial Key Research and Development Program (No. 2019RKE27004), Shandong Provincial Science Foundation for Excellent Young Scholars (No. ZR2020YQ04), Qilu Young Scholar Program of Shandong University, and Taishan Scholar Program of Shandong Province.